# PCL enzymatic hydrolysis: a mechanistic study

*Beatriz C. Almeida,*[a,b] *Pedro Figueiredo* [a,b] *and Alexandra T. P. Carvalho* [*a]

[a] CNC – Center for Neuroscience and Cell Biology, Institute for Interdisciplinary Research (IIIUC), University of Coimbra 3004-504, Coimbra (Portugal)

[b] These authors contributed equally.

*atpcarvalho@uc.pt

**Abstract:** Accumulation of plastic waste is a major environmental problem. Enzymes, particularly esterases, play an important role in the biodegradation of polyesters. These enzymes are usually only active on aliphatic polyesters, but a few have showed catalytic activity for semi-aromatic polyesters. Due to the importance of these processes, an atomic level characterization of how common polyesters are degraded by esterases is necessary. Hereby, we present a Molecular dynamics (MD) and Quantum Mechanics/Molecular Mechanics (QM/MM) MD study of the hydrolysis of a model of polycaprolactone (PCL), one of the most widely used biomaterials, by the thermophilic esterase from the archaeon *Archaeoglobus fulgidus* (AfEST). This enzyme is particularly interesting because it can withstand temperatures well above the glass transition of many polyesters. Our insights about the reaction mechanism are important for the design of customized enzymes able to degrade different synthetic polyesters.

**Introduction:** In the last years, an exponential increase of the environmental problems associated with plastics has stimulated a new interest in aliphatic polyesters for several applications such as degradable materials in packaging, disposables and biomedical materials as alternatives to commodity plastics.[1] These polymers, in particular poly(butylene succinate) - PBS, poly(ε-caprolactone) - PCL, poly(lactic acid) - PLA and poly(3-hydroxybutyrate) - PHB have attracted much interest because of their desirable properties of biocompatibility and biodegradability.[2]

PCL is a bioresorbable polymer with a slow hydrolytic degradation rate (2-3 years) which has been widely used in biomedical applications, such as in nanoparticles for targeted drug delivery systems and in the engineering of tissue scaffolds.[2,3]

Several lipases (EC 3.1.1.3) and some carboxylesterases (EC 3.1.1.1) are active on aliphatic polyesters and thus can be employed in their synthesis or hydrolysis.[4,5] They represent a milder and selective alternative because they do not require the use of toxic reagents, which avoids problems associated with trace residues of metallic catalysts in biomedical applications.[4,6] Compared with conventional chemical routes, enzymatic catalysis gives a more precise construction of well-defined structures, such high control of enantio-selectivity, chemo-selectivity, regio-selectivity, stereo-selectivity and choro-selectivity.[7] These enzymes are characterized by a α/β hydrolase fold and a catalytic triad consisting of an aspartate (Asp) or glutamate (Glu) residue, a histidine (His) and a nucleophilic serine (Ser) residue.[8] They display a ping-pong bi-bi mechanism comprising nucleophilic attack by the catalytic serine, which first or in concerted manner transfers its proton to the catalytic histidine, forming the first tetrahedral intermediate (**INT-1**). This is followed by proton rearrangement and release of the alcohol moiety, which leads to the formation of the acyl-enzyme intermediate (also called enzyme activated monomer - **EAM**). This intermediate is then deacylated by an incoming nucleophile, which is generally a water, an alcohol or an amine. The formation of the second tetrahedral intermediate (**INT-2**) and proton rearrangement finally leads to product release and regeneration of the free enzyme (Scheme 1).[9,10]

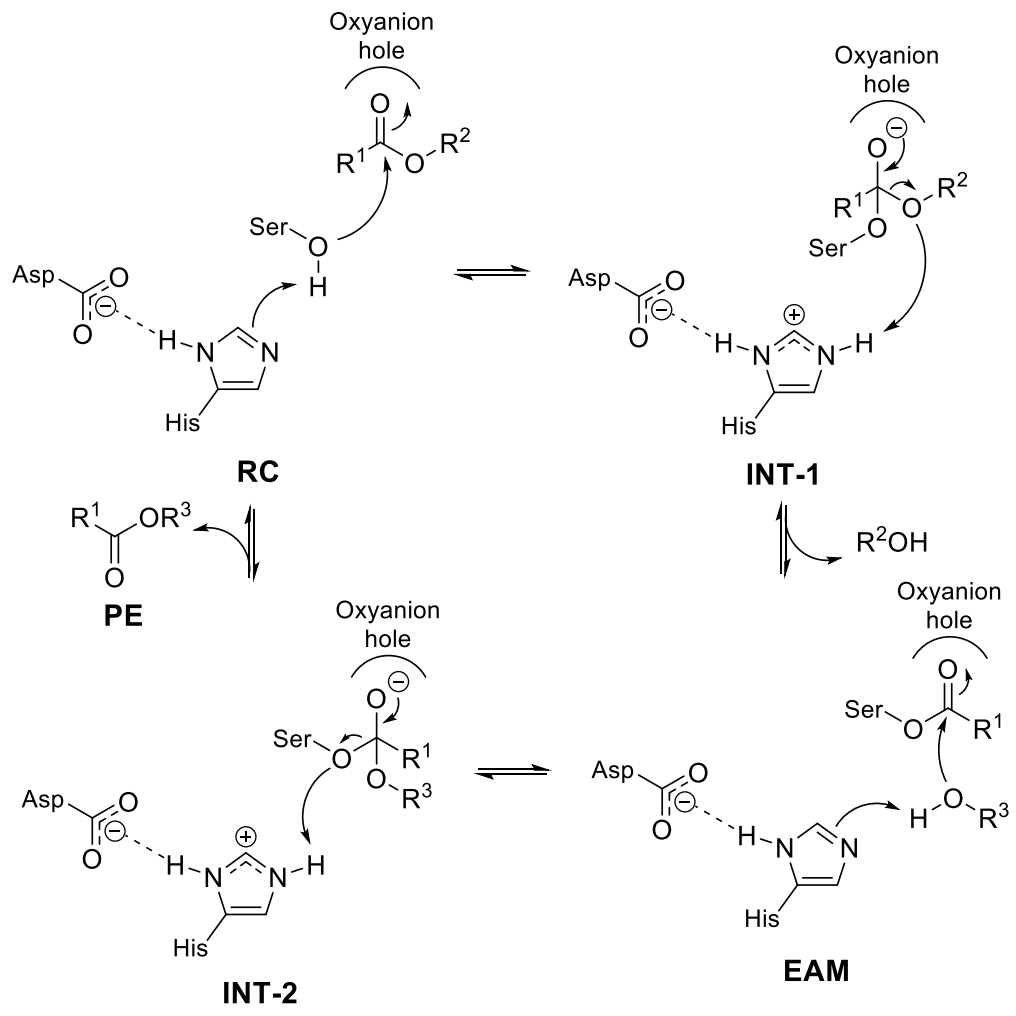

**Scheme 1.** General catalytic cycle for polyesters (**PE**) biosynthesis.

Biosynthesis is achieved via ring-opening polymerization (ROP) of cyclic monomers such as cyclic esters (lactones) and condensation polymerization.[11] In lipase/esterase ROP the first step is the opening of the lactone ring by nucleophilic attack of the serine, forming the **EAM**. Chain initiation occurs through deacylation by water forming 6-hydroxycaproic acid (**6-HCA**) and chain propagation by deacylation by the terminal alcohol of a growing PCL chain. Although, chain initiation is necessary for ROP, **6-HCA** and water are competitive nucleophiles to the growing PCL chain.[12]

Enzymatic hydrolysis is the major factor on the degradation of **PCL** in compost and sea water.[13] The biodegradability of polyesters depends on the chemical and physical properties of the polymer, particularly on the degree of crystallinity.[14] This dependence model was first deduced from the correlation that was found between the difference in the temperature at which the degradation takes place with *Pseudomonas* sp. lipase and the melting point of the polyester.[15] According to this model, polymers with high melting points such as Polyethylene terephthalate (PET) would not be able to undergo biodegradation. Nevertheless, since then several enzymes were discovered that are able to degrade PET. One of those is even active on PET with ≈15% crystallinity.[16]

For a polymer to be hydrolyzed, the polymeric chains need to have enough mobility to reach the enzyme active site. Reactions at higher temperatures can promote higher chain flexibility at the amorphous regions of polymers. Therefore, thermophile enzymes, which allow reactions over harsh conditions without significant loss in enzymatic activity, are promising candidates for potential industrial applications. An example is the esterase from the hyper-thermophilic archaeon *Archaeoglobus fulgidus* (AfEST) that has broad substrate specificity and high stability.[17] This enzyme has been well characterized with concern to the crystal structure, catalytic mechanism and substrate specificity. It has a classical α/β hydrolase fold and a cap domain and displays a dimer arrangement. The cap is composed of five helices from two separate regions (residues 1–54 and 188–246). The catalytic triad $Ser_{160}$-$His_{285}$-$Asp_{255}$ is located between the α/β hydrolase fold and the cap.[8] AfEST was classified as a member of a hormone-sensitive lipase group of the lipase/esterase family, with an optimal temperature of 80 ºC. It can catalyze the hydrolysis of a broad range of esters, including p-nitrophenyl esters, vinyl esters and triacylglycerols with a short chain, which exhibit the highest activity towards p-nitrophenyl hexanoate among the p-nitrophenyl esters tested.[18,19]

The synthesis of PCL catalyzed by AfEST was previously described. The existing forms, free and immobilized, allow the formation of polymeric chains with Molecular mass (Mn) values between 900-1340 g/mol and monomer conversion ratios between 72-99% at 80 ºC.[17,20,21]

Here we have studied the catalytic mechanism for PCL hydrolysis and were able to characterize in detail all intermediates involved in polyester hydrolysis by this enzyme. This study opens the way for future design of tailored enzymes able to hydrolyze a broad range of polyesters.

## Results and Discussion

**Formation of EAM with PCL Model Compound ($PCL_2$):** Analysis of the binding of substrates and intermediates to the AfEST active site, shows that the enzyme displays a medium and a large pocket (Figure 1). Both pockets are located at the interface between the α/β hydrolase fold and the cap domains, but whereas the medium pocket has no opening to the outside of the protein, the large pocket has an opening that allows the access of the hydrolysable ester and nucleophiles to the active site of the enzyme. Consequently, the access to the medium pocket must be through the large pocket (Figure 1). The overall volume of the pockets is 343.454 Å$^3$, which is of considerable size comparing with other esterases/lipases with known structure that are capable to hydrolyze different aliphatic and semi-aromatic polyesters (Table S1).

The oxyanion hole of AfEST is composed by the amide groups of the backbone: $Gly_{88}$, $Gly_{89}$ and $Ala_{161}$. These atoms establish hydrogen bonds with the oxygen atom of the tetrahedral intermediate (Table S2), leading to stabilization of the developing negative charge of this atom during the catalytic cycle. In the reactant complex, the average distance between the oxygen atom of $Ser_{160}$ and the carbon of the carbonyl from substrate that is going to be attacked is 3.82±0.38 Å (Figure 2, Table S2).

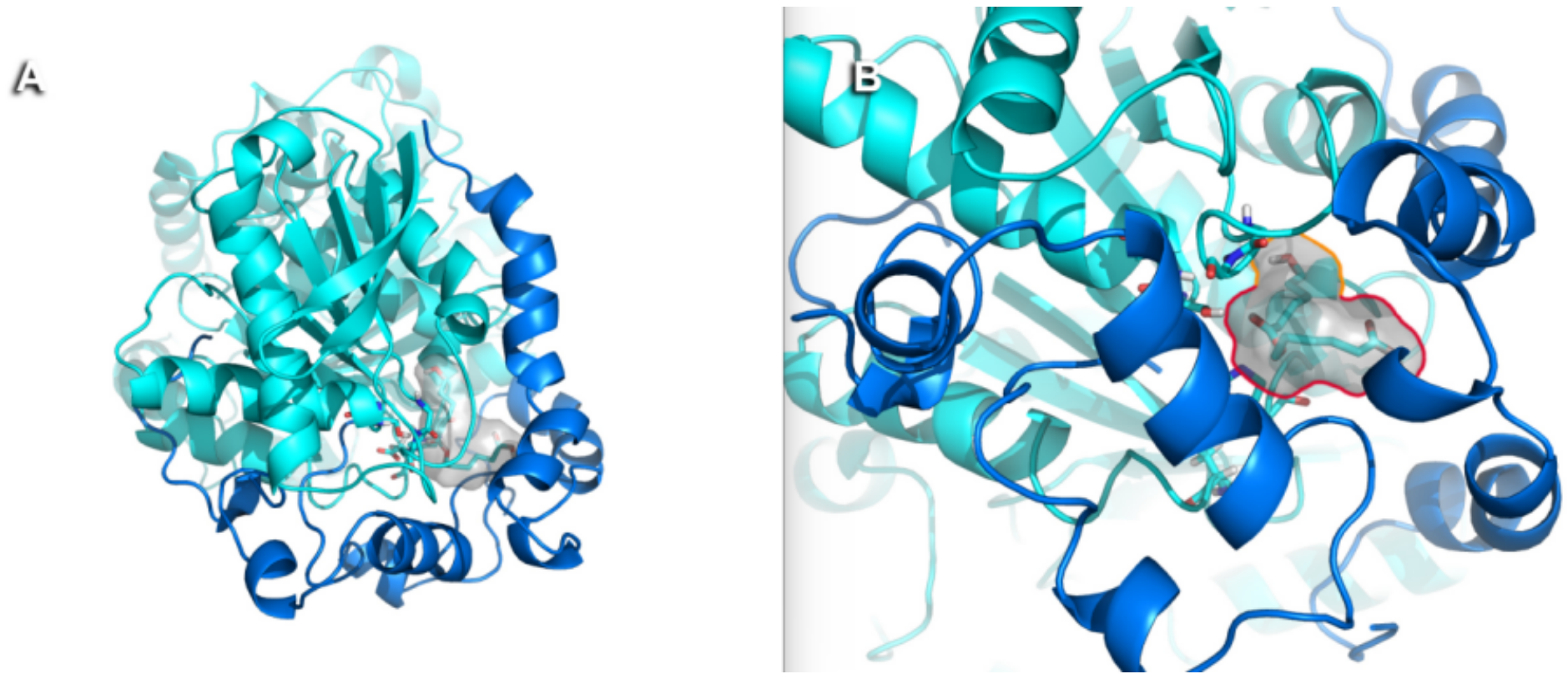

**Figure 1.** A- Structure of AfEST showing the active site at the interface of the α/β hydrolase fold and the cap domains; B- The two pockets accessible to the substrate (medium represented with an orange line and the large with a red line).

The reaction starts with nucleophilic attack performed by the oxygen atom of $Ser_{160}$ on the **PCL** model compound and leading to formation of **INT-1**, which is 3.3 kcal/mol above the reactants (Figure 3, Scheme 2). The transition state connecting the **RC** and **INT-1** structures is $\mathbf{TS_1}$ (Figure S1). In **INT-1** the terminal alcohol function is in the medium pocket and the carbonyl region in the large pocket (Figure 1). The reaction is concerted, meaning that the proton transfer from $Ser_{160}$ to $His_{285}$, which are 3.07±0.58 Å away from each other in the reactants, occurs concomitantly with the formation of the C-O bond between $Ser_{160}$ and the **PCL** model compound (1.14±0.03 Å, Table S2).

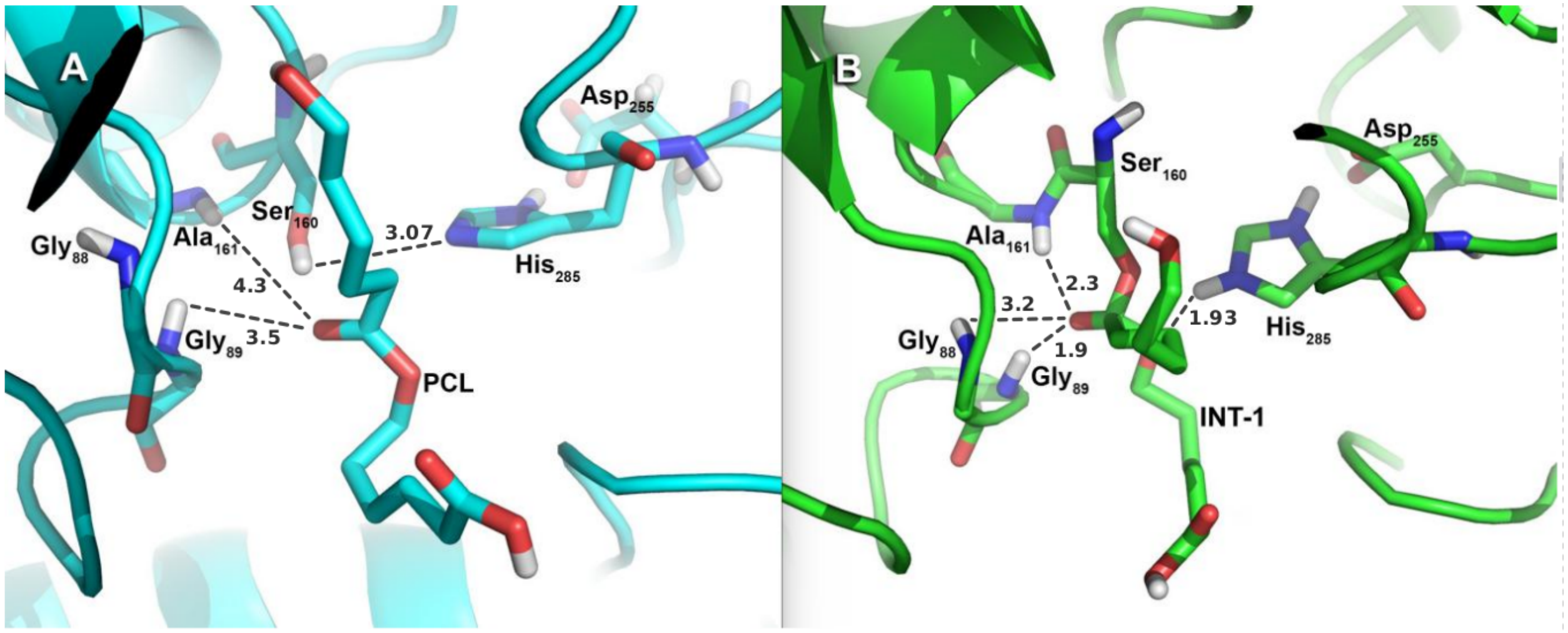

**Figure 2.** A- Structure of the **PCL** model compound at the active site of AfEST; B- Structure of **INT-1**.

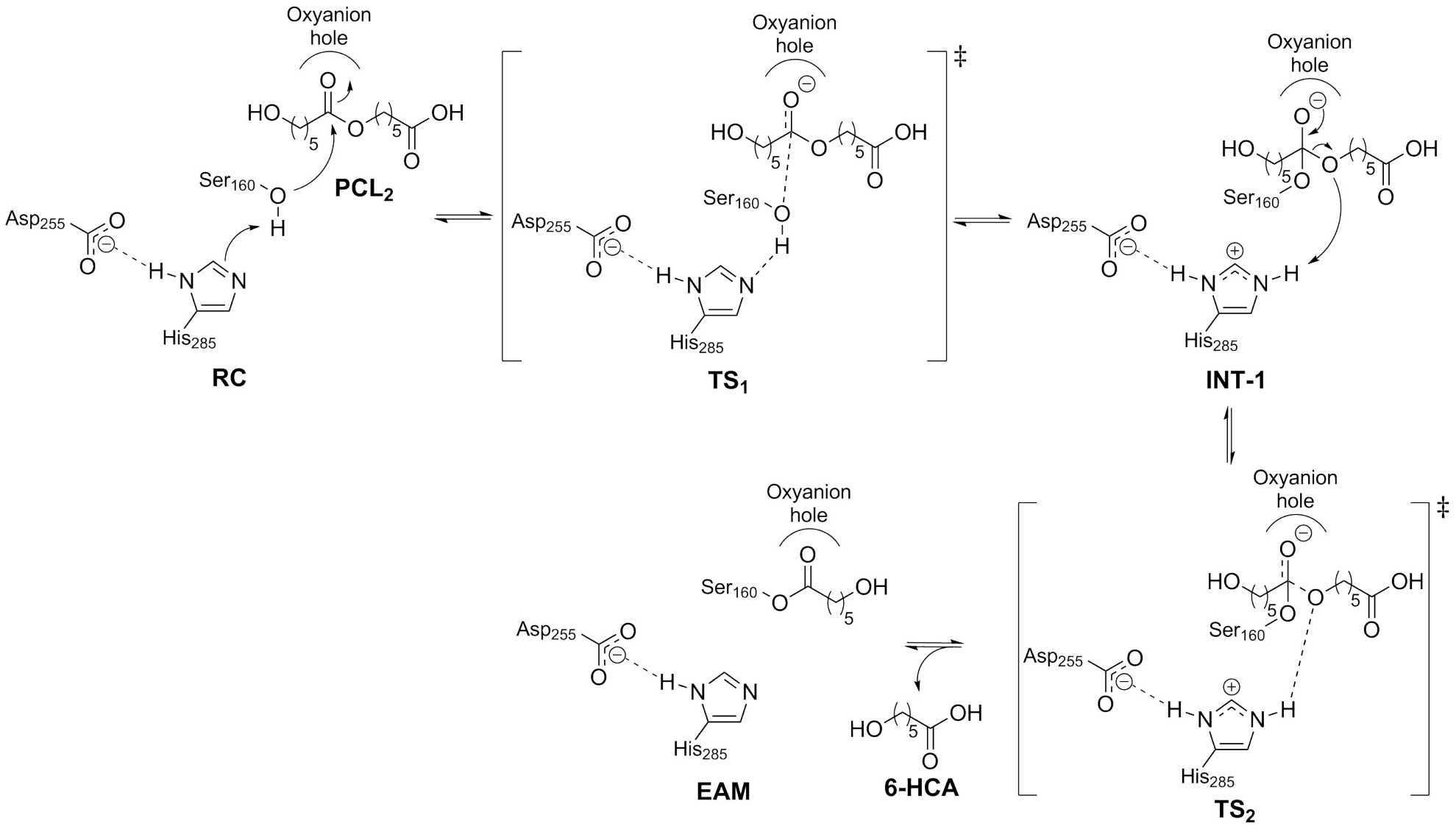

**Scheme 2.** First two steps of the catalytic cycle (from the reactants to the **EAM** intermediate).

In fact, we observe that all the reactions in the cycle (Scheme 2 and 3) comprehend a proton transfer and the formation/ breakage of a C-O bond, occur all in a concerted way. The negatively charged oxygen of **INT-1** is making hydrogen bonds with the amide groups of the oxyanion hole (Figure 2, Table S2).

**INT-1** is converted to the **EAM** with the release of **6-HCA**, being the **TS$_2$** structure (Figure S2) located 9.8 kcal/mol above **INT-1** and the overall $\Delta G^{\ddagger}$ of this step 12.9 kcal/mol (Figure 3).

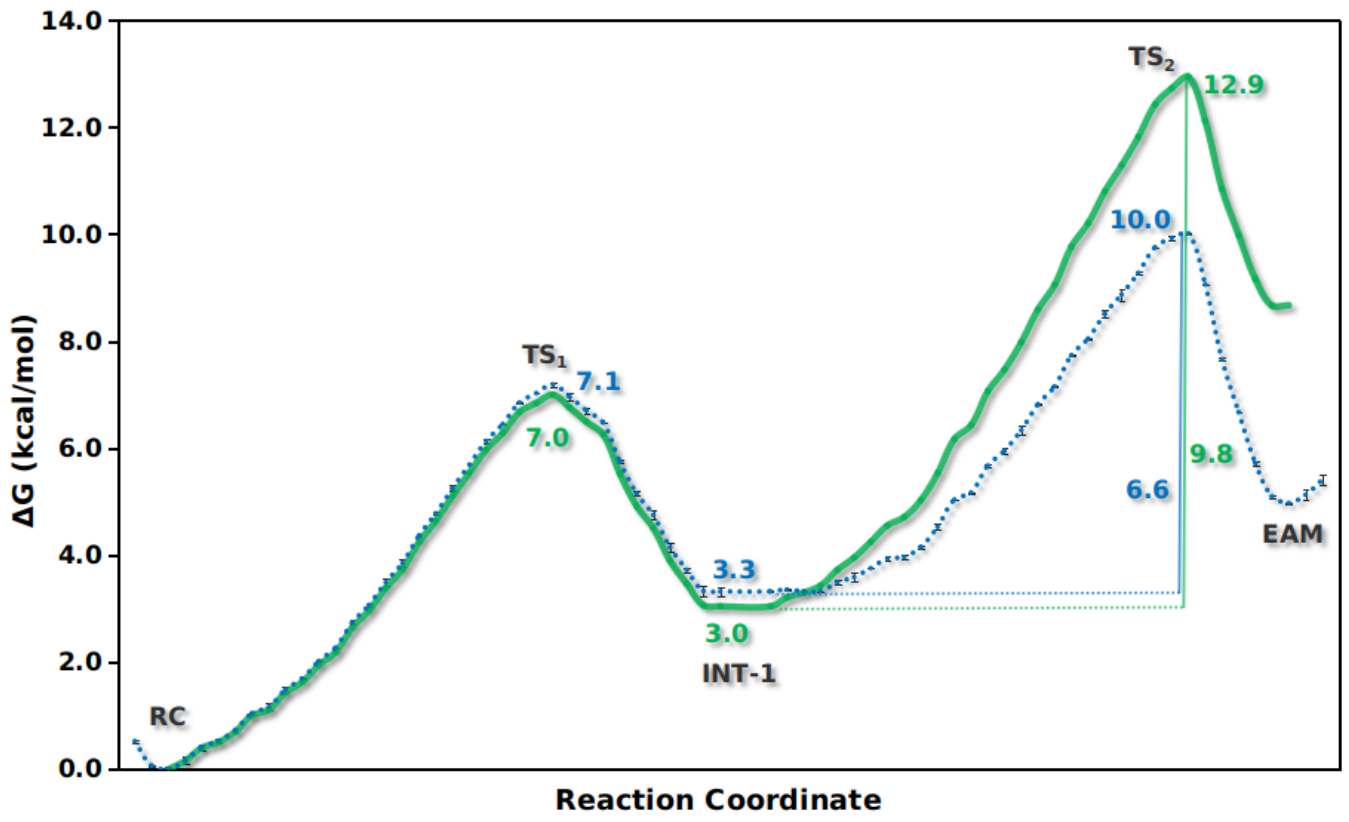

**Figure 3.** Energetic profile for the first half part of the catalytic cycle (formation of intermediate **EAM**). The blue dashed line denotes the energies calculated with the PM6/Amber_parm99SB and the green line corresponds to the corrected energies at the M06-2X/6-31G(d) level.

**Enzyme Deacylation, Formation of 6-Hydroxycaproic Acid:** In the second half part of the catalytic cycle, a water molecule attacks **EAM** generating the **INT-2** via formation of $\mathbf{TS_3}$ (Scheme 3, Figure S3).

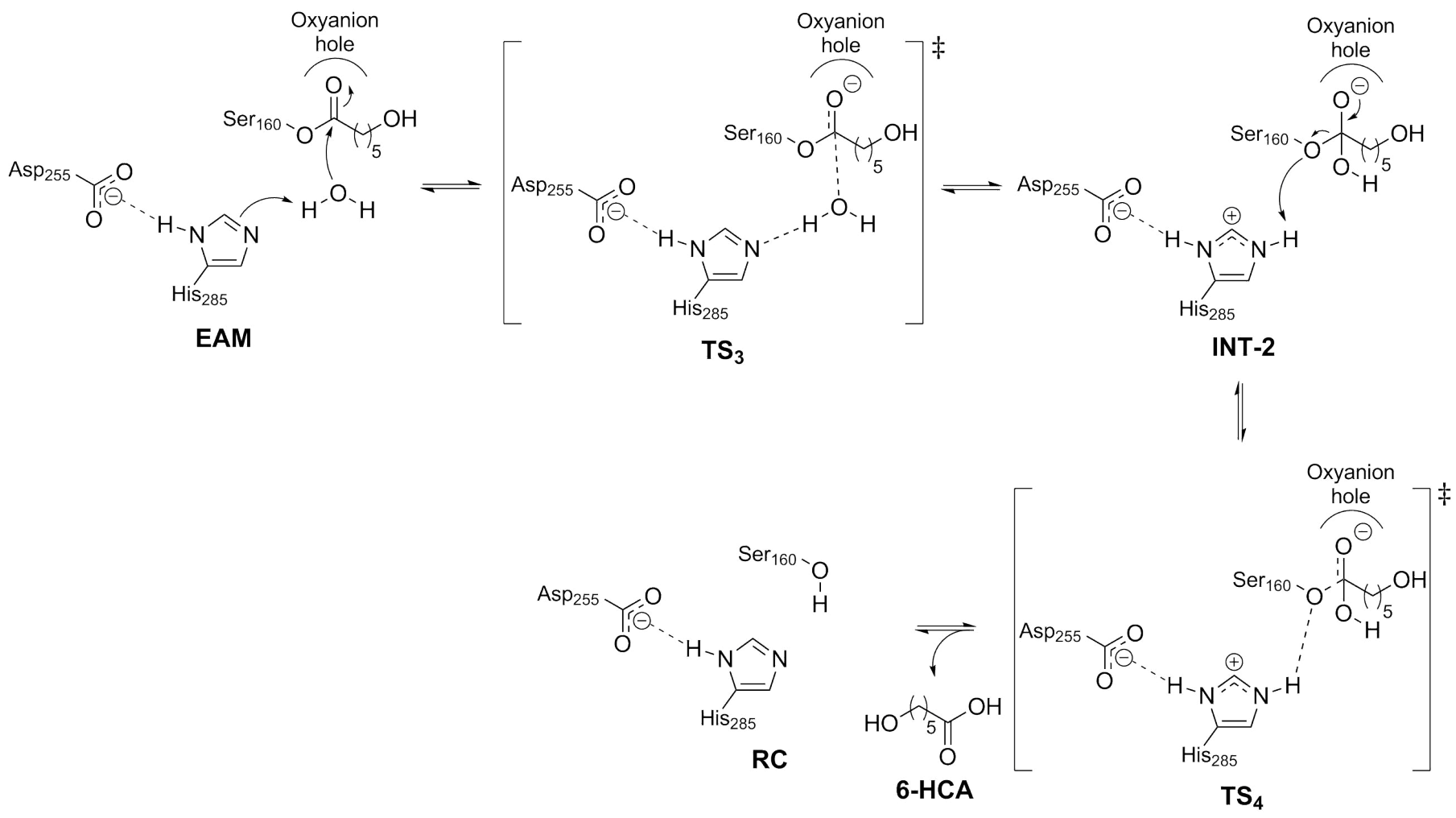

**Scheme 3.** Second half-part of the catalytic cycle of the hydrolysis of $\mathbf{PCL_2}$.

In the **EAM** structure, the alcohol from substrate is in the medium pocket and the average distance of the closest water oxygen atom to the carbon of **EAM** is 6.86 Å. There are several water molecules in the vicinity of the **EAM** (Figure S4). $\mathbf{TS_3}$ is 6.4 kcal/mol above the **EAM**, being the formation of **INT-2** exergonic in relation to the formation of **EAM** with a ΔG of -2.1 kcal/mol (Figure 4). Water will drive the equilibrium towards the acid release and a second molecule of **6-HCA** is released, regenerating the free enzyme that is ready for another turnover. The transition state (**TS$_4$**, Figure S5) associated with this step has a $\Delta G^{\ddagger}$ of 6.6 kcal/mol (Figure 4).

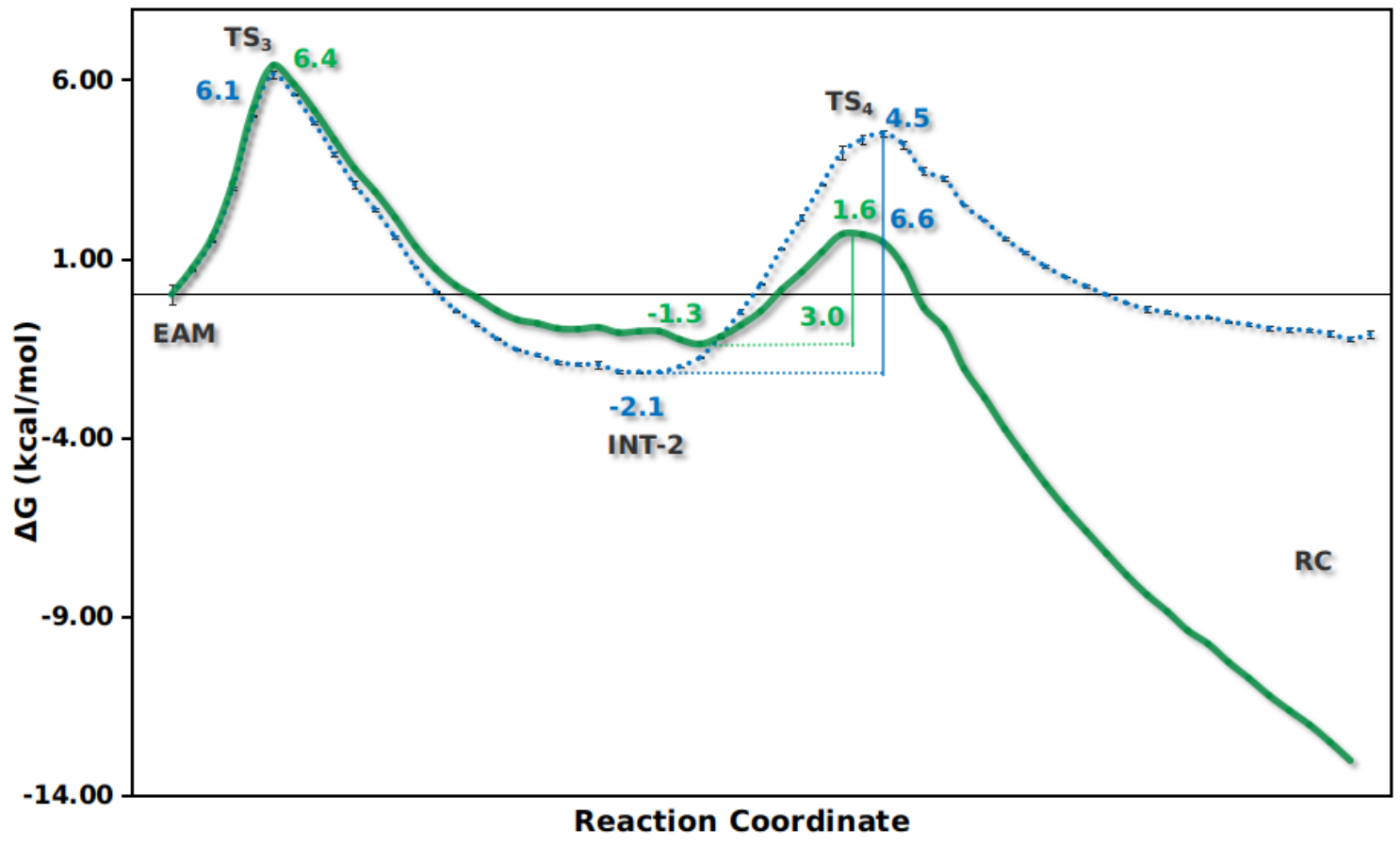

**Figure 4.** Energetic profile for the conversion of the intermediate **EAM** to the **INT-2** and **INT-2** to the products. The blue dashed line denotes the energies calculated with the PM6/Amber_parm99SB and the green line corresponds to the corrected energies at the M06-2X/6-31G(d) level.

In the products, we observe that the carbonyl moiety makes hydrogen bonds with Ser$_{160}$ and His$_{285}$ and the alcohol moiety moves to the large pocket (Figure 5, Table S3).

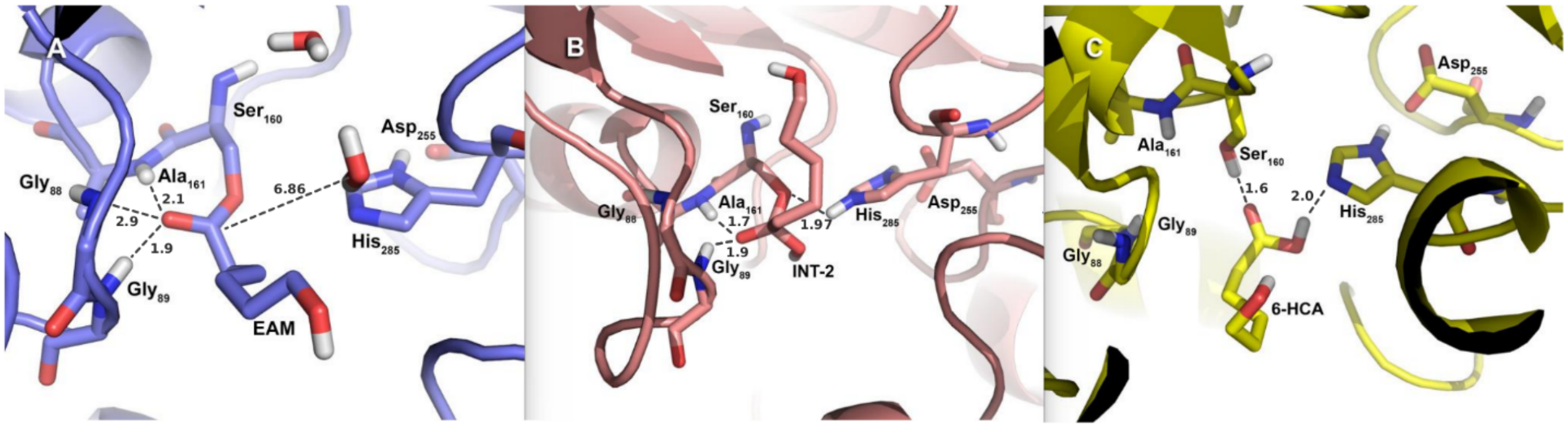

**Figure 5.** A- Structure of the **EAM** complex; B- Structure of the **INT-2** complex; C- Structure of the complex between the free enzyme and **6-HCA** (products).

**Implications for the degradability of polyesters:** AfEST has been extensively characterized in the synthesis of PCL.[17,20,21] It was found that the enzyme can accommodate **PCL** at its active site, however, the low molecular weight (Mw) of the obtained polyesters shows that hydrolysis competes with the synthesis depending on the amount of water available in the active site. We show that there are several water molecules in the vicinity EAM intermediate (Figure S4). The proposed catalytic mechanism shows that the rate-limiting step is easily achievable with a $\Delta G^{\ddagger}$ of around 12.9 kcal/mol (Figure 3). An *in vitro* study of AfEST hydrolysis of PNP hexanoate indicates a barrier of 15.5 kcal/mol, which is expected to be higher than the hydrolysis of **PCL** due to the aromatic moiety of this substrate.[19] Furthermore, other carboxylesterases were found to promote **PCL** depolymerization.[22] Comparison of the AfEST active site with other esterases/lipases with known structure also shows that AfEST has a large combined volume of the active site pockets (Figure S3).

The physical and chemical properties of the polymer have a strong influence on enzymatic hydrolysis. **PCL** has a low melting point of 60 ºC and a glass-transition temperature ($T_g$) of -60 ºC. However, for polyesters with higher melting points, the degree of crystallinity is a major limiting factor for hydrolysis. Nevertheless, the amorphous region will contain polymers chains that can access the active site and it was suggested that as the amorphous regions are degraded, the crystalline regions can be ultimately degraded.[23] Consequently, thermophilic enzymes able to operate at higher temperatures than the $T_g$ of many types of polyester are good candidates for protein engineering and can help solving the plastic pollution problems that we are currently facing.

**Conclusions:** We have conducted a QM/MM MD study of the hydrolysis of a model compound of PCL by the thermophilic esterase from the hyper-thermophilic archaeon *Archaeoglobus fulgidus*. Our results show that the hydrolysis has an overall $\Delta G^{\ddagger}$ of just 12.9 kcal/mol, showing that the chemistry step is already very efficient. This makes this enzyme a good starting point for further rational engineering efforts to improve polymer binding while keeping the high thermostability.

## Experimental Section

**Modelling:** The initial structure was modelled from AfEST crystal structure pdb code 1JJI.[8] Hydrogen atoms were added according to the protonation states of the residues taking in account a medium with pH 7 (the enzyme has optimal activity for pH 6.5 to 7.5).[19]

The reactants, products and intermediates (**INT-1**, **INT-2** and **EAM**) were geometry optimized in Gaussian09[24] using B3LYP[25] with the 6-31G(d) basis set and a Polarizable Continuum Model (PCM)[26] solvent description. Atomic partial charges were calculated resorting to the Restrained Electrostatic Potential (RESP)[27] method from HF/6-31G(d) single point energy calculations.

**Molecular Docking:** Molecular docking was performed using the AutoDock 4.2 suite of programs with the Lamarckian genetic algorithm (LGA)[28]. A grid box was centered on the oxygen of the side chain of the catalytic serine (residue 160 for AfEST). A total of 100 LGA runs were carried out for each ligand-protein complex. The population was 300, the maximum number of generations was 27,000 and the maximum number of energy evaluations was 2,500,000.

**Molecular Dynamics:** MD simulations were performed using the Amber molecular dynamics program (AMBER18)[29] with the parm99SB[30] and GAFF[31] force fields. The structures were placed within an octahedral box of TIP3P[29] waters (the distance between the protein surface and the box was set to 10 Å) and counter ions were added to make the entire system neutral. The systems were subjected to two initial energy minimizations with the steepest descent and conjugate gradient algorithms and to 500 ps of equilibration in a NVT ensemble using Langevin dynamics with small restraints on the protein (10 kcal/mol) to heat the system from 0 K to 300 K. Production simulations were carried out at 300 K in the NPT ensemble using Langevin dynamics with a collision frequency of 1.0 $ps^{-1}$. Constant pressure periodic boundary conditions were imposed with an average pressure of 1 atm. Isotropic position scaling was used to maintain pressure with a relaxation time of 2 ps. The time step was set to 2 fs. SHAKE constraints were applied to all bonds involving hydrogen atoms.[32] The particle mesh Ewald (PME) method[33] was used to calculate electrostatic interactions with a cut-off distance of 10 Å. All the minima in the catalytic cycle were subjected to 50 ns triplicate simulations with different initial velocities, for a total combined time of 750 ns. Reference structures, corresponding to the lowest root-mean-square deviation (RMSD) structures to the average of the simulations, were calculated.[34]

**Quantum Mechanical/Molecular Dynamics (QM/MM) Calculations:** The QM/MM MD calculations[35] were performed using the internal semi-empirical hybrid QM/MM functionality implemented in AMBER18 with periodic boundary conditions. The PM6[36] semi-empirical method was employed for the high-level layer and the MM region was described by the Amber parm99SB force field.[30] The choice of the PM6 Hamiltonian represents a good compromise between the use of a fast and a reliable QM method and the extensive sampling required to accurately determine enzymatic barriers.[36,37]

Electrostatic embedding[38] was employed and the boundary was treated via the hydrogen link atom approach. Long-range electrostatic interactions were described by an adapted implementation of the PME method for QM/MM.[39]

The high-level layer in the reactants complex included the **PCL** model compound, $Ser_{160}$, the side chains of $His_{285}$ and $Asp_{255}$, and the amide groups of $Gly_{88}$, $Gly_{89}$ and $Ala_{161}$ (Figure S6). The total number of atoms in the high-level layer in the reactants was 78. For the other intermediates, the high-level layer included the same protein residues plus either the **INT-1**, the **EAM**, the **INT-2** or the product **6-HCA**. The reaction coordinates used in **RC** to **INT-1** and **INT-1** to **EAM** were the distance between the $PCL_2(C_a)$ - $Ser_{160}(O_b)$, and $Ser_{160}(O_g)$ - $His_{285}(H_c)$, respectively. Concerning to the second part of the catalytic cycle, the reaction coordinate for **EAM** to **INT-2** was the distance between $Ser_{160}(O_h)$ - $His_{285}(H_c)$ and for **INT-2** to **RC** the distance between 6-HCA($C_a$) - $Ser_{160}(O_b)$.

The initial structures were **INT-1** and **INT-2** and the reaction coordinates were restrained in 0.1 Å steps using the umbrella sampling method, except near the transition states were smaller steps of 0.02 Å were employed. The potentials of mean force (PMFs) were computed resorting to the Weighted Histogram Analysis Method (WHAM) method.[40]

Corrections were applied to the obtained PM6 PMFs by performing geometry optimizations of the high-level layer models with the exchange correlation functional M06-2X[41] and the 6-31G(d) basis set according to Carvalho *et al*. and Bownan *et al*..[42,43] We provide in the Supporting information the coordinates of these structures (Tables S5-S10).

**Supporting Information:** The comparison of the total active site volume for esterases/lipases that accept aliphatic and semi-aromatic polyesters(Table S1); The average distance between atoms in Schemes 2 and 3 (Tables S2 and S3); The PMFs with the corrections calculated with B3LYP with dispersion corrections, M06-2X and wB97XD (Figure S7); Coordinate lists for all structures and respective method (Tables S5-S10).

Representation of the QM region, as well of the transition states TS1-TS4 (Figures S1, S2, S3, S5 and S6); The number of water molecules in first and second solvation shells of EAM intermediate (Figure S4);

**Acknowledgements:** We thank Fundação para a Ciência e a Tecnologia for the financial support through the exploratory project MIT-Portugal (MIT-EXPL/ISF/0021/2017) and the grant IF/01272/2015.

**Authors Orcid**

Beatriz C. Almeida: 0000-0002-1278-803X

Pedro Figueiredo: 0000-0002-1243-0265

Alexandra Carvalho*: 0000-0003-2827-5527

**References**

**(1)** Bednarek, M. Branched aliphatic polyesters by ring-opening (co)polymerization. *Prog. Polym. Sci. 2016, 58, 27–58.*

**(2)** Kobayashi, S. Enzymatic ring□opening polymerization and polycondensation for the green synthesis of polyesters. *Polym. Adv. Technol.* **2015**, 26, 677–686.

**(3)** Abedalwafa, M.; Wang, F.; Wang, L.; Li, C. Biodegradable poly-epsilon-caprolactone (PCL) for tissue engineering applications: a review. *Rev. Adv. Mater. Sci.* **2013**, 34, 123-140.

**(4)** Kadokawa, J.; Kobayashi, S. Polymer synthesis by enzymatic catalysis. Curr. Opin. *Chem. Biol.* **2010**, 14, 145–153.

**(5)** Cao, H.; Han, H.; Li, G.; Yang, J.; Zhang, L.; Yang, Y.; Fang, X.; Li, Q. Biocatalytic synthesis of poly (δ-valerolactone) using a thermophilic esterase from *Archaeoglobus fulgidus* as catalyst. *Int. J. Mol. Sci.* **2012**, 13, 12232–12241.

**(6)** Zhang, J.; Shi, H.; Wu, D.; Xing, Z.; Zhang, A.; Yang, Y.; Li, Q. Recent developments in lipase-catalyzed synthesis of polymeric materials. *Process Biochem.* **2014**, 49, 797–806.

**(7)** Veld M. A. J.; Palmans, A. R. A. Enzymatic Polymerisation. *Adv. Polym. Sci.*, Springer-Verlag Berlin Heidelberg, Berlin, Heidelberg, **2011**, 55–78.

**(8)** De Simone, G.; Menchise, V.; Manco, G.; Mandrich, L.; Sorrentino, N.; Lang, D.; Rossi, M.; Pedone, C. The crystal structure of a hyper-thermophilic carboxylesterase from the archaeon *Archaeoglobus fulgidus1. J. Mol. Biol.* **2001**, 314, 507-518.

**(9)** Bezborodov, A. M.; Zagustina, N. A. Lipases in catalytic reactions of organic chemistry. *Appl Biochem Microbiol* **2014**, 50, 313–337.

**(10)** Raza, S.; Fransson, L.; Hult, K. Enantioselectivity in *Candida antarctica* lipase B: A molecular dynamics study. *Protein Sci.* **2001**, 10, 329–338.

**(11)** Labet, M.; Thielemans, W. Synthesis of polycaprolactone: a review. *Chem. Soc. Rev.* **2009**, 38, 3484–3504.

**(12)** Shirahama, H.; Shiomi, M.; Sakane, M.; Yasuda, H. Biodegradation of Novel Optically Active Polyesters Synthesized by Copolymerization of (R)-MOHEL with Lactones. *Macromolecules* **1996**, 29, 4821–4828.

**(13)** Krasowska, K.; Heimowska A.; Rutkowska, M. Enzymatic and hydrolytic degradation of poly (ε-caprolactone) in natural conditions. *Int. Polym. Sci. Technol.* **2006**, 33, 57-62.

**(14)** Mueller, R.-J. Biological degradation of synthetic polyesters—enzymes as potential catalysts for polyester recycling. *Process Biochem.* **2006**, 41, 2124–2128.

**(15)** Marten, E.; Müller, R.-J.; Deckwer, W.-D. Studies on the enzymatic hydrolysis of polyesters I. Low molecular mass model esters and aliphatic polyesters. *Polym. Degrad. Stab.* **2003**, 80, 485–501.

**(16)** Austin, H. P.; Allen, M. D.; Donohoe, B. S.; Rorrer, N. A.; Kearns, F. L.; Silveira, R. L.; Pollard, B. C.; Dominick, G.; Duman, R.; Omari, K. E.; Mykhaylyk, V.; Wagner, A.; Michener, W. E.; Amore, A.; Skaf, M. S.; Crowley, M. F.; Thorne, A. W.; Johnson, C. W.; Woodcock, H. L.; McGeehan, J. E.; Beckham, G. T. Characterization and engineering of a plastic-degrading aromatic polyesterase. *PNAS* **2018**, 115, 4350–4357.

**(17)** Ma, J.; Li, Q.; Song, B.; Liu, D.; Zheng, B.; Zhang, Z.; Feng, Y. Ring-opening polymerization of □-caprolactone catalyzed by a novel thermophilic esterase from the archaeon *Archaeoglobus fulgidus*. *J. Mol. Catal. B: Enzym.* **2009**, 56, 151–157.

**(18)** D'Auria, S.; Herman, P., Lakowicz, J. R.; Bertoli, E.; Tanfani, F.; Rossi M.; Manco, G. The Thermophilic Esterase from *Archaeoglobus fulgidus*: Structure and Conformational Dynamics at High Temperature. *Proteins: Struct., Funct., Genet.* **2000**, 38, 351–360.

**(19)** Manco, G.; Giosuè, E.; D'Auria, S.; Herman, P.; Carrea, G.; Rossi, M. Cloning, Overexpression, and Properties of a New Thermophilic and Thermostable Esterase with Sequence Similarity to Hormone-Sensitive Lipase Subfamily from the Archaeon *Archaeoglobus fulgidus*. *Arch. Biochem. Biophys.* 2000, 373, 182–192.

**(20)** Ren, H.; Xing, Z.; Yang, J.; Jiang, W.; Zhang, G.; Tang, J.; Li, Q. Construction of an Immobilized Thermophilic Esterase on Epoxy Support for Poly (ε-caprolactone) Synthesis. *Molecules* **2016**, 21, 796.

**(21)** Li, G.; Li, Q. Increasing the productivity of TNFR-Fc in GS-CHO cells at reduced culture temperatures. *Biotechnol. Bioprocess Eng.* **2011**, 16, 1201–1207.

**(22)** Hajighasemi, M.; Nocek, B. P.; Tchigvintsev, A.; Brown, G.; Flick, R.; Xu, X.; Cui, H.; Hai, T.; Joachimiak, A.; Golyshin, P. N.; Savchenko, A.; Edwards E. A.; Yakunin, A. F. Biochemical and Structural Insights into Enzymatic Depolymerization of Polylactic Acid and Other Polyesters by Microbial Carboxylesterases. *Biomacromolecules*, **2016**, 17, 2027–2039.

**(23)** Mochizuki, M.; Hirano, M.; Kanmuri, Y.; Kudo K.; Tokiwa, Y. Hydrolysis of polycaprolactone fibers by lipase: Effects of draw ratio on enzymatic degradation. *J. Appl. Polym. Sci.*, **1995**, 55, 289–296.

**(24)** Frisch, M.; Trucks, G.; Schlegel, H.; Scuseria, G. ; Robb, M. ; Cheeseman, J. ; Scalmani, G. ; Barone, V.; Mennucci, B.; Petersson, G.; Nakatsuji, H.; Caricato, M.; Li, X.; Hratchian, H.; Izmaylov, A.; Bloino, J.; Zheng, G.; Sonnenberg, J.; Hada, M.; Ehara, M.; Toyota, K.; Fukuda, R.; Hasegawa, J.; Ishida, M.; Nakajima, T.; Honda, Y.; Kitao, O.; Nakai, H.; Vreven, T.; Montgomery, J.; Peralta, J.; Ogliaro, F.; Bearpark, M.; Heyd, J.; Brothers, E.; Kudin, K.; Staroverov, V.; Kobayashi, R.; Normand, J.; Raghavachari, K.; Rendell, A.; Burant, J.; Iyengar, S.; Tomasi, J.; Cossi, M.; Rega, N.; Millam, J.; Klene, M.; Knox, J.; Cross, J.; Bakken, V.; Adamo, C.; Jaramillo, J.; Gomperts, R.; Stratmann, R.; Yazyev, O.; Austin, A.; Cammi, R.; Pomelli, C.; Ochterski, J.; Martin, R.; Morokuma, K.; Zakrzewski, V.; Voth, G.;

Salvador, P.; Dannenberg, J.; Dapprich, S.; Daniels, A.; Foresman, J.; Ortiz, J.; Cioslowski J.; Fox, D. Gaussian 09, Revision B.01, Gaussian, Inc., Wallingford CT.

**(25)** Ashvar, C. S.; Devlin, F. J.; Bak, K. L.; Taylor P. R.; Stephens, P. J.; Ab Initio Calculation of Vibrational Absorption and Circular Dichroism Spectra: 6,8-dioxabicyclo[3.2.1]octane. *J. Phys. Chem.*, **1996**, 100, 9262–9270.

**(26)** Tomasi, J.; Mennucci B.; Cammi, R. Quantum Mechanical Continuum Solvation Models. *Chem. Rev.*, **2005**, 105, 2999–3094.

**(27)** Bayly, C. I.; Cieplak, P.; Cornell, W.; Kollman, P. A. A well-behaved electrostatic potential based method using charge restraints for deriving atomic charges: the RESP model. *J. Phys. Chem.*, **1993**, 97, 10269–10280.

**(28)** Morris, G. M.; Huey, R.; Lindstrom, W.; Sanner, M. F.; Belew, R. K.; Goodsell D. S.; Olson, A. J. AutoDock4 and AutoDockTools4: Automated docking with selective receptor flexibility. *J. Comput. Chem.*, **2009**, 30, 2785–2791.

**(29)** Salomon□Ferrer, R.; Case D. A.; Walker, R. C. An overview of the Amber biomolecular simulation package. *Wiley Interdiscip. Rev. Comput. Mol. Sci.*, **2013**, 3, 198–210.

**(30)** Hornak, V.; Abel, R.; Okur, A.; Strockbine, B.; Roitberg A.; Simmerling, C. Comparison of multiple Amber force fields and development of improved protein backbone parameters. *Proteins: Struct., Funct., Bioinf.*, **2006**, 65, 712–725.

**(31)** Wang, J.; Wolf, R. M.; Caldwell, J. W.; Kollman, P. A.; Case, D. A. Development and testing of a general amber force field. *J. Comput. Chem.*, **2004**, 25, 1157–1174.

**(32)** Ryckaert, J. P.; Ciccotti, G.; Berendsen, H. J. C. Numerical integration of the cartesian equations of motion of a system with constraints: molecular dynamics of n-alkanes. *J. Comput. Phys.*, **1977**, 23, 327–341.

**(33)** Darden, T.; York D.; Pedersen, L. Particle mesh Ewald: An N-log(N) method for Ewald sums in large systems. *J. Chem. Phys.*, **1993**, 98, 10089-10092.

**(34)** Dourado, D. F. A. R.; Swart, M.; Carvalho, A. T. P. Why the Flavin Adenine Dinucleotide (FAD) Cofactor Needs to be Covalently Linked to Complex II of the Electron□Transport Chain for the Conversion of FADH2 into FAD. *Chem. Eur. J.*, **2018**, 24, 5246–5252.

**(35)** Carvalho, A. T. P.; Barrozo, A.; Doron, D.; Kilshtain, A. V.; Major, D. T.; Kamerlin, S. C. L. Challenges in computational studies of enzyme structure, function and dynamics. *J. Mol. Graph. Model.*, **2014**, 54, 62–79.

**(36)** Stewart, J. J. P. Optimization of parameters for semiempirical methods V: Modification of NDDO approximations and application to 70 elements. *J. Mol. Model.*, **2007**, 13, 1173–1213.

**(37)** Jindal, G.; Warshel, A.; Exploring the Dependence of QM/MM Calculations of Enzyme Catalysis on the Size of the QM Region. *J. Phys. Chem. B*, **2016**, 120, 9913–9921.

**(38)** Bakowies, D.; Thiel, W. Hybrid Models for Combined Quantum Mechanical and Molecular Mechanical Approaches. *J. Phys. Chem.*, **1996**, 100, 10580–10594.

**(39)** Nam, K.; Gao, J.; York, D. M. An Efficient Linear-Scaling Ewald Method for Long-Range Electrostatic Interactions in Combined QM/MM Calculations. *J. Chem. Theory Comput.*, **2005**, 1, 2–13.

**(40)** Grossfield A. "WHAM: The Weighted Histogram Analysis Method", http://membrane.urmc.rochester.edu/content/wham.

**(41)** Zhao, Y.; Truhlar, D. G. The M06 Suite of Density Functionals for Main Group Thermochemistry, Thermochemical Kinetics, Noncovalent Interactions, Excited States, and Transition Elements: Two New Functionals and Systematic Testing of Four M06-Class Functionals and 12 Other Functionals. Theor. Chem. Acc. 2008, 120 (1), 215–241.

**(42)** Carvalho, A. T. P.; Dourado, D. F. A. R.; Skvortsov, T.; Abreu, M.; Ferguson, L. J.; Quinn, D. J.; Moodyb, T. S.; Huang, M. Catalytic mechanism of phenylacetone monooxygenases for non-native linear substrates. *Phys. Chem. Chem. Phys.*, **2017**, 19, 26851–26861.

**(43)** Bowman, A. L.; Grant, I. M.; Mulholland, A. J. QM/MM simulations predict a covalent intermediate in the hen egg white lysozyme reaction with its natural substrate. *Chem. Commun.*, **2008**, 37, 4425–4427.